\documentclass[conference]{IEEEtran}
\IEEEoverridecommandlockouts
\usepackage{cite}
\usepackage{amsmath,amssymb,amsfonts}
\usepackage[indLines=false, italicComments=false, commentColor=black]{algpseudocodex}
\usepackage{graphicx}
\usepackage{textcomp}
\usepackage{xcolor}
\usepackage{enumitem}
\usepackage{multirow}
\usepackage{algorithm}
\usepackage{caption}
\usepackage{subcaption}
\usepackage{chngcntr}
\def\BibTeX{{\rm B\kern-.05em{\sc i\kern-.025em b}\kern-.08em
    T\kern-.1667em\lower.7ex\hbox{E}\kern-.125emX}}

\algnewcommand\algorithmicforeach{\textbf{for each}}
\algdef{S}[FOR]{ForEach}[1]{\algorithmicforeach\ #1\ \algorithmicdo}

\counterwithout{algorithm}{section}

\begin{document}

\title{Building a Graph-based Deep Learning network model from captured traffic traces\\

\thanks{This publication is part of the Spanish I+D+i project TRAINER-A (ref.PID2020-118011GB-C21), funded by MCIN/ AEI/10.13039/501100011033. This work is also partially funded by the Catalan Institution for Research and Advanced Studies (ICREA), and by the Joan Oró predoctoral program, from the Secretariat for Universities and Research, part of the Ministry of Research and Universities of the Goverment of Catalonia, and the European Social Fund Plus (ref. BDNS 657443).}
}

\author{\IEEEauthorblockN{1\textsuperscript{st} Carlos Güemes Palau}
\IEEEauthorblockA{\textit{Barcelona Neural Networking Center} \\
\textit{Universitat Politècnica de Catalunya}\\
Barcelona, Spain \\
carlos.guemes@upc.edu}
\and
\IEEEauthorblockN{2\textsuperscript{nd} Miquel Ferriol Galmés}
\IEEEauthorblockA{\textit{Barcelona Neural Networking Center} \\
\textit{Universitat Politècnica de Catalunya}\\
Barcelona, Spain \\
miquel.ferriol@upc.edu}
\and
\IEEEauthorblockN{3\textsuperscript{rd} Albert Cabellos Aparicio}
\IEEEauthorblockA{\textit{Barcelona Neural Networking Center} \\
\textit{Universitat Politècnica de Catalunya}\\
Barcelona, Spain \\
albert.cabellos@upc.edu}
\and
\IEEEauthorblockN{4\textsuperscript{th} Pere Barlet Ros}
\IEEEauthorblockA{\textit{Barcelona Neural Networking Center} \\
\textit{Universitat Politècnica de Catalunya}\\
Barcelona, Spain \\
pere.barlet@upc.edu}
}

\maketitle

\begin{abstract}
Currently the state of the art network models are based or depend on Discrete Event Simulation (DES). While DES is highly accurate, it is also computationally costly and cumbersome to parallelize, making it unpractical to simulate high performance networks. Additionally, simulated scenarios fail to capture all of the complexities present in real network scenarios. While there exists network models based on Machine Learning (ML) techniques to minimize these issues, these models are also trained with simulated data and hence vulnerable to the same pitfalls. Consequently, the Graph Neural Networking Challenge 2023 introduces a dataset of captured traffic traces that can be used to build a ML-based network model without these limitations. In this paper we propose a Graph Neural Network (GNN)-based solution specifically designed to better capture the complexities of real network scenarios. This is done through a novel encoding method to capture information from the sequence of captured packets, and an improved message passing algorithm to better represent the dependencies present in physical networks. We show that the proposed solution it is able to learn and generalize to unseen captured network scenarios.
\end{abstract}

\begin{IEEEkeywords}
network modelling, neural networks, graph neural networks, recurrent neural networks \end{IEEEkeywords}

\section{Introduction}
\label{sec:intro}
In recent years, network modeling has risen in prominence as one of the most active research fields related to computer networks. Correctly designed network models can be used to simulate network configurations without risk, as it does not involve using the actual network, and try out scenarios that may be too rare or risky to encounter in real life. Arguably, the most prevalent way to build these network models is through the use of discrete event simulation (DES) methodologies. Notable examples include the ns-3~\cite{Riley2010} and OMNeT++~\cite{Varga2019} simulators. However, while DES-based simulators tend to be accurate, they also tend to be slow and scale poorly with the simulation scenario size (e.g., amount of traffic, size of the network...) \cite{JAFER201354}. Also, simulation tends to be bounded by only considering idealized scenarios in which all variables and patterns are known, which usually it is not the case in real life (e.g. the traffic distribution and its parameters \cite{6036768, 392383, Popoola2017EmpiricalPO}).

Hence, as of late, there has been a push to complement DES with Machine Learning (ML) methodologies. These hybrid techniques offer similar accuracy while offering a speed up in performance by replacing part of the simulation with ML models~\cite{10.1145/3452296.3472926, 10.1145/3544216.3544248}. Alternatively, there are end-to-end ML models that entirely replace DES simulators~\cite{9109574, ferriolgalmés2022routenetfermi}. These tend to be Graph Neural Networks (GNNs) \cite{4700287}, as it is a type of neural network (NN) that is able to exploit relational information and therefore able to understand the inter-dependencies between the different devices in the network, independently of its topology and size. Still, the ML models present in these solutions are trained with DES-generated data, resulting in them inheriting the same constraints and challenges associated with DES simulators.

\begin{figure}[t]
    \centering
    \includegraphics[width=\linewidth]{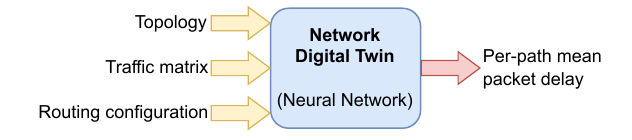}
    \caption{Scheme of the neural network-based solution requested in the Graph Neural Networking Challenge}
    \label{fig:challenge_solution}
\end{figure}

The objective behind the 2023 edition of the Graph Neural Network Challenge is the creation of a network model that does not share the same limitations as DES simulators. This means the creation of a network model purely from captured data, rather than using DES-simulators or DES-generated data. To do so, we have built a testbed server that is able to simulate different network topologies under different levels of load and two traffic flow distributions. The result is a dataset formed from 8604 samples, each sample representing one traffic scenario in the testbed. Participants of the challenge are asked to design GNN-based models using this dataset, as summarized in Figure \ref{fig:challenge_solution}.

In this paper, we present one possible solution for the Graph Neural Network Challenge 2023. Our proposal is the continuation of the RouteNet family of models, in which we introduce two novel adaptations for dealing with captured data. On one hand, we introduce a new way of encoding the network traffic distribution through the use of a Recurrent Neural Network (RNN), a type of neural network meant to deal with sequences of data. On the other, we also updated the GNNs message passing paradigm to be better suited for the more complex real networks. The results show that with these changes, our model achieves significant improvements in modeling packet delays of captured traffic data. The remainder of the paper is structured as follows:
\begin{itemize}
    \item In section \ref{sec:background} we briefly introduce GNNs and RNNs and their notations.
    \item In section \ref{sec:solution}, we introduce the proposed solution to the Graph Neural Network Challenge 2023.
    \item In section \ref{sec:evaluation}, we evaluate the proposed solution relative to a baseline model based on the previous generation of the RouteNet models, RouteNet-Fermi \cite{ferriolgalmés2022routenetfermi}.
\end{itemize}

\section{Background}
\label{sec:background}
\subsection{Graph Neural Networks}
Relational information can be expressed in the form of a graph $G \in \{V, E\}$, itself identified by a set of vertices $v \in V$ connected by a set of edges $e \in E$. Hence, GNNs \cite{4700287} are the family of NNs designed precisely to work with and exploit relational information. They have been successfully applied in many fields in which relational information is available, which also includes network modeling \cite{10.1145/3477141, 9846958}.

There are plenty of GNN variants, each defined with a different mechanism to extract and utilize relational information \cite{10.1145/3477141}. For example, spectral GNNs update the encoding for each vertex through a convolution operation based on the spectrum of the Laplacian matrix \cite{bruna2014spectral}, while spatial GNNs define its convolution using the vertex's neighborhood in the original topology \cite{4773279}. Independently of the specific GNN architecture used, the fact that its mechanism is equivariant to vertex and edge permutations grants GNNs \textit{strong relational inductive bias}, that is, the ability to generalize across topologies \cite{battaglia2018relational}.

Specifically, our proposed solution is based on Message-Passing NN (MPNN) \cite{10.5555/3305381.3305512}. They are defined as a more general framework of spatial GNNs which identify their models in three phases: an initial encoding phase where the initial representation for each vertex is obtained, followed by a message passing phase in which neighboring vertices share information, and ending in the readout phase in which the vertices' states are used to generate the final output. Specifically, the message-passing phase is defined as the following steps:
\begin{enumerate}
    \item Message generation: each node $v \in V$ generates a message to its neighbors $w \in N_v, N_v \subseteq V $ using both nodes' states $h_v$ and $h_w$, and edge information $e_{(v,w)}$:
    $$ m_{w \rightarrow v} = M(h_v, h_w, e_{(v,w)}) \hspace{5mm} \text{iff } w \in N_v $$
    \item Message aggregation: each node aggregates the messages from its neighbors:
    $$ m_v = agg(\{m_{w \rightarrow v}, w \in N_v \}) $$
    \item Update: the aggregated messages are used to update the current node state:
    $$ h_v = U(h_v, m_v) $$
\end{enumerate}
This process is repeated for a fixed number of iterations. Both functions $M$ and $U$ are learnable, differentiable functions that can be implemented by NNs. The aggregation operator can be any commutative operation, such as the sum of all messages, as (generally) neighboring vertices are not described by an order relation.

\subsection{Recurrent Neural Networks}

RNNs \cite{Rumelhart1986} are a type of state-full NNs. Unlike most NNs, which can be represented as a pure mathematical function $y = f(x)$, whose output $y$ is only affected by its input $x$, RNNs also utilize a state $h$. This makes RNNs very useful for dealing with data structured as a sequence, such as time series. Also, unlike other types of models (e.g. transformers), it can also accept input sequences of undetermined length, which makes them highly versatile. To use an RNN, besides the input $x$ the RNN also requires the current state $h$. The RNN will not only produce the output $y$, but also a modified state $h'$:
$$ y, h' = RNN(x; h) $$
In order to process a sequence $X = [x_0, ..., x_n]$, it can be thought of as the RNN iterates through every single element to return the output sequence $Y$:
\[
Y, h' = RNN(X; h) \Leftrightarrow 
\begin{array}{l}
y_0, h' = RNN(x_0; h) \\
\textbf{for } i \in [1, n] \textbf{ do:} \\
\hspace{3mm} y_i, h' = RNN(x_i; h')
\end{array}
\]

Nowadays the most commonly used architectures for RNN are the Long Short-Term Memory (LSTM) \cite{10.1162/neco.1997.9.8.1735} and the more recent Gated Recurrent Unit (GRU) \cite{cho2014properties} architectures, which offers similar expressive power as the former with fewer parameters \cite{chung2014empirical}. In our solution, we use RNNs in instances where the information is structured as sequences, such as when encoding packet traffic or when processing the evolution of a traffic flow as it crosses through the network.

\section{Our proposed solution}
\label{sec:solution}

In this section, we introduce our proposed solution for the Challenge, a GNN-based solution designed to handle the difficulties of real network data. To do so, the model includes two improvements over its predecessor. First, the message-passing algorithm has been expanded to capture the complexities of real network scenarios. Second, our proposed solution also includes using an RNN-based encoder that can process complex, non-parameterized traffic distributions through sequences of captured packets.\\

\begin{figure}[t]
    \centering
    \includegraphics[width=\linewidth]{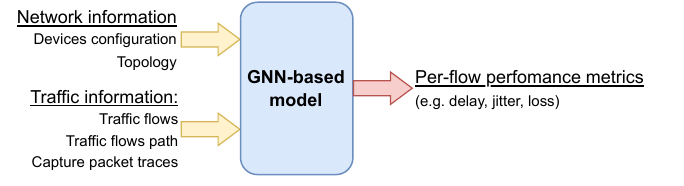}
    \caption{Black-box scheme of the proposed GNN model}
    \label{fig:black_box}
\end{figure}

Figure \ref{fig:black_box} shows a black-box representation of this model. The model takes as input the characteristics of the network scenario. These characteristics can be divided into two groups:
\begin{itemize}
    \item Characteristics of the underlying network: its devices, including links, routers, and switches; and its topology.
    \item Characteristics of the traffic flows: their source, their destination, their path across the network, and their size both in terms of the number of bits transmitted and the number of packets. We also include sequences of captured packets, that will later be used to learn about the underlying traffic distributions.
\end{itemize}
The output of the model focuses on obtaining per-flow performance metrics, although in the context of the challenge and this paper, we will focus on obtaining the per-flow mean delay.

Ultimately, the objective of the model is to exploit an accurate representation of the network components to generalize to unseen topologies and scenarios, while capturing the complexities of captured traffic data. In this section, we will begin by explaining how we formally represent network components and their relationships. We will then follow by explaining how the initial representations for the network elements are obtained. At the end of the section, we will summarize our proposed solution.

\subsection{Representing network components and dynamics}
\label{sec:network_dynamics}

First, let us define a physical network as a set of devices $\mathcal{D} \in \{d_i : i \in (1, ..., n_d)\}$ connected through a set of links $\mathcal{L} = \{l_i : i \in (1, ..., n_l)\}$. Each device $d_i$ is composed by a set of ports $\mathcal{Q} = \{q_i : i \in (1, ..., n_q)\}$, such as $d_i \subset \mathcal{Q}$. Physical links and ports have a 1-to-1 relationship, that is, each physical link is connected to a unique port, and each port is connected to a unique link. In the context of the challenge's networks, these devices can either be switches or routers.

A network is also populated by a set of flows $\mathcal{F} =  \{f_i : i \in (1, ..., n_f)\}$. Each flow is defined by the path of links and ports that it traverses through the network, forming a sequence of tuples of ports and links, defined as $f_i = \{(q_{p_q(f_i,0)}, l_{p_l(f_i, 0)}), ..., (q_{p_q(f_i,|f_i|)}, l_{p_l(f_i, |f_i|)})\}$, where $p_q$ and $p_l$ return the index of the $j$-th port or link along the path of flow $f_i$, respectively. For commodity later on, we will also define the function $Q_f(q_i)$, which returns all the flows that go through port $q_i$ and its associated link, and the function $Q_d(q_i)$ which returns the device which the port $q_i$ belongs to. 

From the network elements we have defined, we can build a graph that represents the relationships between links, ports, devices, and flows. However, we must also represent the following network dynamics:
\begin{enumerate}[label=(\roman*)]
    \item Each link is associated with a single port, each port is associated to a single link.
    \item The state of ports of the same device will be affected by each other due to their dependencies on shared resources (e.g., a device's shared memory and computing power).
    \item The state of flows (e.g. throughput, losses) is affected by the state of the links and ports they traverse.
    \item The state of links and ports (e.g. link utilization, port queue utilization) will depend on the state of flows passing through them.
\end{enumerate}

Now, we present how these properties can be formalized:

\subsubsection*{(i) Each link is associated to a single port, each port is associated to a single link} the most straightforward way to represent this property is to encode each port-link pair together with $h_{lq_i}$, where $h_{lq_i}$ is the encoded state for the port-link pair $lq_i \equiv (l_i, q_i)$. While another solution would have been to define a set of functions that maps between the two, this would have increased the complexity of our proposed solution without necessarily increasing its expressive power.

\subsubsection*{(ii) The state of ports of the same device will be affected by each other due to their dependencies on shared resources}
\[
h_{d_i} = G_d(\{ h_{lq_j}, q_j \in d_i \}) \label{eq:G_d} \tag{1}
\]
Where $h_{d_i}$ is the encoded state of the device $d_i$, and $G_d$ is an unknown function that takes as input all the states belonging to the ports of $d_i$ and outputs its encoded state.

\subsubsection*{(iii) The state of flows is affected by the state of the links and ports they traverse}
\[ h_{f_i} = G_f([h_{Q_d(q_{f_i(1)})}, h_{lq_{f_i(1)}}, ..., h_{Q_d(q_{f_i(|f_i|)})}, h_{lq_{f_i(|f_i|)}}]) \label{eq:G_f} \tag{2} \]
Where $h_{f_i}$ is the encoded state of the flow $f_i$, and $G_f$ is an unknown function that takes as input the sequence of states belonging to the devices, links, and ports that flow $f_i$ traverses and outputs its encoded state. Specifically, for the $j$-th position in $f_i$'s path, $h_{Q_d(q_{f_i(j)})}$ represents the state of the device which the $j$-port belongs to, and $h_{lq_{f_i(j)}}$ represents the state of the $j$-link-port pair.

\subsubsection*{(iv) The state of links and ports will depend on the state of flows passing through them}
\[ h_{lq_i} = G_{lq}(\{h_{f_j}, f_j \in Q_f(q_i)\}) \label{eq:G_lq} \tag{3} \]
Where $G_{lq}$ is an unknown function that takes as input all the encoded states from flows that go through the link-port $lq_i$ and returns the encoded state for it.

Note that these properties show circular dependencies between the different states of the elements in the network. As a result, we will use an iterative algorithm that starts with the initial set of encoded states $h^0_{lq}, h^0_{f}, h^0_{d}$, and in each iteration they are refined until convergence. This will be the message-passing algorithm, although its description will be introduced later in subsection \ref{sec:algorithm}.

\subsection{Extracting initial embedding}
\label{sec:initial_embedding}

However, before we can run the iterative algorithm to refine the encoded states, we need to obtain the initial embeddings for each of the network elements. These embeddings are defined using the elements' characteristics within the network scenarios.

\begin{table}[h]
\begin{tabular}{|l|l|l|}
\hline
\textbf{Network Device} & \textbf{Features}                             & \textbf{Notes}            \\ \hline
Link information        & Bandwidth                                     & In bits per second        \\ \hline
Device information      & Device type                                   & Router or Switch            \\ \hline
\multirow{5}{*}{Flow}   & Average load                                  & In bits per second        \\ \cline{2-3} 
                        & Number of packets                             &                           \\ \cline{2-3} 
                        & Packet size                                   & In bits                   \\ \cline{2-3} 
                        & \multirow{2}{*}{\textit{Sequence of packets}} & Features per packet:      \\
                        &                                               & timestamp and packet size \\ \hline
\end{tabular} \vspace{1mm}
\caption{Extracted features of each network element for initial embeddings}
\label{tab:features}
\end{table}

The extracted features per network device can be seen in Table \ref{tab:features}. Note that there may be other relevant features that we have not included. For example, features like the flow's quality of service or the link's propagation delay have not been included as they are constant across the entire dataset, but may be included if needed. Other features, such as each port's buffer memory size, have not been included since it is not information that is usually available to the network operator, and therefore unrealistic to be added to the network model.

Out of the identified features, the link bandwidth, and the flow's average load, packet size and number of packets have all been normalized with min-max normalization. The device type is a categorical feature with two possible values. For our solution have decided to separate ports and device encoding functions into two separate functions, one for each type of device. This grants us more expressive power than encoding the device type through one-hot encoding. While it scales poorly to the number of devices, this is not an issue as in Challenge's dataset only considers two types of devices.

To encode the link and port features, the link and device features are used with the encoding function $E_{lq}$, to obtain the initial embeddings:
\[
h^0_{lq_i} = E_{lq}(\{\text{Link bandwidth}, \text{Port's device type}\}) 
\]
\[ = \begin{cases} E_{lq, router}(\{\text{Link bandwidth}\}) & \text{if $q_i$ belongs to router} \\ E_{lq, switch}(\{\text{Link bandwidth}\}) & \text{if $q_i$ belongs to switch} \end{cases} \label{eq:E_lq} \tag{4}
\]
Then, to obtain the device embeddings, all the port embedding belongings to the same device are aggregated through the sum operator followed by the encoding function $E_d$:
\[
h^0_{d_i} = E_{d}(sum(\{ h_{lq_j}, q_j \in d_i \})) 
\]
\[
= \begin{cases} E_{d, router}(sum(\{ h_{lq_j}, q_j \in d_i \})) & \text{if $d_i$ is router} \\ E_{d, switch}(sum(\{ h_{lq_j}, q_j \in d_i \})) & \text{if $d_i$ is switch} \end{cases} \label{eq:E_d} \tag{5}
\]

To encode the traffic flows, we first need to encode their sequences of packets. To do so, we apply the following process for each flow $f_i$:
\begin{enumerate}
    \item First we trim the sequence to only include the first second of captured traffic. We then aggregate the sequence, where we obtain the number packets sent per millisecond. As in the Challenge's dataset the packet size in each flow is constant, this is equivalent to the number of bits transmitted. This reduces the length of the sequence significantly, which makes it faster and easier to handle by the model later on.
    \item We train a sequential encoder model $E_{pkts}$ to process the entire sequence into a single encoded vector $h_{pkts_i}$.
    \[
    h_{pkts_i} = E_{pkts}(agg(\{\text{Sequence of packets}\}))
    \]
    \item We input the encoded vector with the average load, number of packets, and average packet size into an encoder function $E_f$ to obtain the initial flow embeddings.
    \[
    h^0_{f_i} = E_f(h_{pkts_i} | \{\text{Avg. load, Avg. packet size, Num. packets}\}) \label{eq:E_f} \tag{6}
    \]
\end{enumerate}

\subsection{Summary of the model}

\begin{algorithm}[!t]
\caption{Summary of our solution}
\label{alg:summary}
\begin{algorithmic}[1]
\Require {$ \mathcal{F}, \mathcal{Q}, \mathcal{L}, \mathcal{D}, \boldsymbol{x}_f, \boldsymbol{x}_{f, pkts}, \boldsymbol{x}_q, \boldsymbol{x}_l $}
\Ensure {$ \boldsymbol{\hat{y}}_f $}
\LComment{Initial encoding phase}
\ForAll {$f \in \mathcal{F}$} \Comment{\footnotesize Initial Flow Embeddings} 
    \State $\boldsymbol{h}_{f, pkts} = \boldsymbol{0}_{pkts}$ \Comment{\footnotesize Encoder RNN initial state is a zero vector}
    \ForAll {$x_{f, pkt} \in \boldsymbol{x}_{f, pkts}$}
        \State $\emptyset, \boldsymbol{h}_{f, pkts} = RNN_{pkts}(x_{f, pkt}; {h}_{f, pkts})$ \EndFor
    \State $\boldsymbol{h}^0_f \gets E_f(\boldsymbol{x_f}|\boldsymbol{h}_{f, pkts}) $
\EndFor
\State $ \mathcal{LQ} \equiv \{(l, q), l \in L \land q \in Q \land paired(l, q) \} $ \Comment{\footnotesize Set of paired links and ports} 
\ForAll {$(l, q) \in \mathcal{LQ} $} $\boldsymbol{h}^0_{lq} \gets E_{lq} (\boldsymbol{x_l}|\boldsymbol{x_q})$ \Comment{\footnotesize Initial Link and Port Embeddings} \EndFor 
\ForAll {$d \in D $} \Comment{\footnotesize Initial Device Embeddings}
    \State $M^0_d \gets \sum_{(l,q) \in \mathcal{LQ} \land q \in d} \boldsymbol{h}^0_{lq}$
    \State $\boldsymbol{h}^0_{d} \gets E_d(M^0_d)$
\EndFor

\LComment{Message Passing Phase}
\State $t = 0$
\Repeat
    \State $t \gets t+1$
    \ForAll {$f \in \mathcal{F}$} \Comment{\footnotesize Message Passing on Flows}
        \State $pos \gets 0$ \Comment{\footnotesize Curr. position in path}
        \State $\boldsymbol{h}^t_f \gets \boldsymbol{h}^{(t-1)}_f$
        \ForAll {$(l,q) \in f$}
            \State $\widetilde{m}^t_{f, pos}, \boldsymbol{h}^t_f \gets RNN_{flows}(\boldsymbol{h}^{t-1}_{lq}; \boldsymbol{h}^t_f)$
            \State $pos \gets pos + 1$
        \EndFor
    \EndFor
    \State $\widetilde{m}^t \gets \{ \widetilde{m}^t_f, f \in \mathcal{F} \}$
    \ForAll {$(l, q) \in \mathcal{LQ}$} \Comment{\footnotesize MP on Links and Ports}
        \State $M^t_{lq} \gets \sum_{(f, pos) \in \hat{Q}_f(l,q)}{\widetilde{m}^t_{(f, pos)}}$ \Comment{\footnotesize L\&Q aggregation}
        \State $\boldsymbol{h}^t_{lq}, \emptyset \gets U_q(M^t_{lq}; \boldsymbol{h}^{t-1}_{lq})$ \Comment{\footnotesize L\&Q update}
        \State $\widetilde{m}^t_{lq} \gets \boldsymbol{h}^t_{lq}$ \Comment{\footnotesize L\&Q message generation}
    \EndFor
    \ForAll{$d \in \mathcal{D}$} \Comment{\footnotesize Message Passing on Devices}
        \State $M^t_{d} \gets \sum_{(l,q) \in \mathcal{LQ} \land q \in d}{\widetilde{m}^t_{lq}}$ \Comment{\footnotesize Device aggregation}
        \State $\boldsymbol{h}^t_d, \emptyset \gets U_q(M^t_{d}; \boldsymbol{h}^{t-1}_d)$ \Comment{\footnotesize Device update, msg. generation}
    \EndFor
\Until{$t > T \lor \widetilde{m}^t \simeq \widetilde{m}^{t-1}  $}

\LComment{Readout Phase}
\ForAll{ $f \in \mathcal{F}$}
    \State $\hat{d}_q \gets \sum_{\widetilde{m}^t_{f, pos} \in \widetilde{m}^t_f} R(\widetilde{m}^t_{f, pos})$ \Comment{\footnotesize Queuing delay}
    \State $\hat{d}_t \gets \sum_{(l,q) \in f} \boldsymbol{x}_{f_{ps}} / \boldsymbol{x}_{l_c}$ \Comment{\footnotesize Transmission delay}
    \State $\hat{d}_{p} \gets \sum_{(l,q) \in f} \boldsymbol{x}_{l_{prop}}$ \Comment{\footnotesize Propagation delay}
    \State $\boldsymbol{\hat{y}}_{f_d} \gets \hat{d}_q + \hat{d}_t + \hat{d}_p$
\EndFor
\end{algorithmic}
\end{algorithm}

The summary of the solution is described at Algorithm \ref{alg:summary}. To do so, our model can be divided into three phases, following the MPNN paradigm:
\begin{enumerate}
    \item Initial encoding (lines 1-15): initial embeddings are built from the network and traffic flows's features.
    \item Message passing (lines 17-34): the different embeddings are enriched through our multi-stage message passing algorithm which represents the network's dynamics.
    \item Readout (lines 36-40): the final traffic embeddings are used in this phase to obtain the final packet delay.
\end{enumerate}

The initial encoding phase is implemented as described back in subsection \ref{sec:initial_embedding}. Specifically, lines 3-7 are dedicated to obtaining the flow embeddings, lines 9-11 the link and port embeddings, and lines 13-15 the device embeddings. The encoding functions for flows $E_f$, links and ports $E_{lq}$, and devices $E_{d}$, are a single Multilayer Perceptron (MLP) each. The sequential model used to encode the packet sequences (lines 4-6) is a two-layered RNN (i.e., equivalent to two RNNs where the output sequence of the first is used as the input sequence of the second one) using GRU cells. To do so, we introduce the entire pre-processed packet sequence, and we retrieve the final internal state from the second layer.

After building the initial representations the model begins the message-passing phase. This is an iterative process and exploits the properties described back in subsection \ref{sec:network_dynamics}. At each iteration, the embeddings are updated as follows:
\begin{enumerate}
    \item The flow embedding (lines 20-26): Here we use a GRU-cell RNN model $RNN_{flows}$ to learn function $G_f$ as described back in Equation (\ref{eq:G_f}). $RNN_{flows}$ takes the sequence of internal states from links, ports, and devices, and returns the sequences of partial flow state after each point in the sequence $\widetilde{m}^t_f$. The final internal state of the RNN acts as the new flow embedding $h^t_f$. As such, the entire process with $RNN_{flows}$ includes the aggregation, update, and message generation steps of the message-passing paradigm.
    \item The link and port embedding (lines 27-30): These lines replicate the behavior described back in Equation (\ref{eq:G_lq}). For each flow, the adequate partial flow states $\widetilde{m}^t_{f, pos}$ are retrieved and aggregated through addition (line 28). To correctly retrieve the partial flow states we defined function $\hat{Q}_f(l,q)$, which given a link-port pair $(l,q)$ as input it returns the set of flows traversing through them, as well as their position in the flow's path.
    The update of the link and port state $h^t_{lq}$ is done through the function $U_q$, implemented as a GRU cell, which takes as input the aggregated messages and as initial state the previous iteration's embedding $h^{t-1}_{lq}$ (line 29).
    \item The device embedding (lines 31-33): this step is nearly identical to updating the link and port embedding, while representing the property described in Equation (\ref{eq:G_d}).
\end{enumerate}
The iterative process runs for a maximum of $T$ iterations, or until the sequences of partial flow embeddings $\widetilde{m}^t$, which will be later used, have converged (line 34). We determine convergence if the mean-absolute relative difference between $\widetilde{m}^t$ and $\widetilde{m}^{t-1}$ is lower than 5\% in 95\% of flows in the scenario. These values are hyperparameters of the solution.

Finally, in the readout phase, the final mean packet delay $\hat{y}_d$ is computed for each flow. The delay is divided into the queuing delay, transmission delay, and propagation delay, which are later added together (line 40). The transmission delay (line 38) and propagation delay (line 39) can be computed directly from the flow features, as we know the average flow bandwidth, link capacity, and link propagation delay. For computing the queuing delay (line 37), each of the flow partial states is passed through the readout function $R$ to obtain the marginal queuing delay attributed at each link-port pair, and then added. Function $R$ is modeled through an MLP.

\label{sec:algorithm}

\section{Evaluation}
\label{sec:evaluation}
In the following section, we cover how we evaluated our solution. Specifically, we first introduce the testbed and the characteristics of the generated dataset. Then, we compare how accurate the proposed solution is when predicting the mean flow delay compared to a baseline model, itself inspired by a naive adaptation of RouteNet-Fermi \cite{ferriolgalmés2022routenetfermi} before introducing the improvements discussed in Section \ref{sec:solution}.

\subsection{Testbed environment}

\begin{figure}[b]
    \centering
    \includegraphics[width=0.9\linewidth]{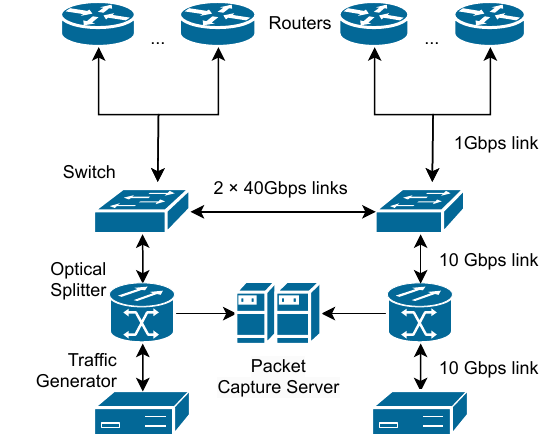}
    \caption{Diagram summarizing the testbed's structure}
    \label{fig:testbed}
\end{figure}

To obtain captured traffic traces needed for the Challenge, a testbed network was built to first generate them. Figure \ref{fig:testbed} shows the structure of the testbed and its components. How the testbed works is summarized as follows:
\begin{itemize}
    \item The testbed can be formed with up to 8 Huawei NetEngine 8000 M1A routers connected to one of the two Huawei S5732-H48UM 2CC 5G Bundle switches.
    \item The switches act as the backbone of the testbed, connecting the routers, the traffic generators and packet capture servers to each other.
    \item Two TRex traffic generators send traffic in such a way that it passes through the indicated routers before ending in either of the traffic generators.
    \item Traffic between the traffic generators and the switches is copied using an optical splitter and then fed to the packet capture server using the PF\_RING software.
\end{itemize}

The traffic generators can generate two traffic distributions:
\begin{itemize}
    \item Constant Bit Rate (CBR): short, regular, high-frequency bursts that average out to a specified flow rate.
    \item Multi-Burst (MB): similar to CBR, but rather than transmitting the entire flow at once, traffic is characterized by periods of no activity interrupted by fixed-length bursts (in terms of number of packets).
\end{itemize}

The configuration based on switches allows for the testbed to vary the topology of the network with ease. In each network scenario, a topology is determined and fixed, and multiple flows are generated. Scenarios can be split into two groups: scenarios that include only MB flows, and scenarios that include both CBR and MB flows. In the first group, there can only be one flow per source-destination pair (as in the first and last router in the path). In the second group, there can be up to one CBR flow and one MB flow per source-destination pair. Also, if two flows share the same source-destination pair they also share the same routing path.

Scenarios lasted for 10 seconds, but only packets sent in the last 5 seconds were captured. This is due that at the beginning of the scenario, the network remains in a transient state, where flows are just starting, buffers are still being filled, and the flow rates are still too irregular. Also, due to the cost of storing all the packet traces and simulating the scenarios, there is a limit to the amount of samples we can generate. Table \ref{tab:samples} shows how many samples were used for training, validation, and testing, and how many of which were used for each scenario.

\begin{table}[h!]
\centering
\begin{tabular}{|l|l|l|l|}
\hline
\textbf{Scenario type}  & \textbf{Training} & \textbf{Validation} & \textbf{Test} \\ \hline
\textbf{CBR+MB}         & 3372              & 843                 & 150           \\ \hline
\textbf{MB}             & 3512              & 877                 & 150           \\ \hline
\textit{\textbf{Total}} & \textit{6884}     & \textit{1720}       & \textit{300}  \\ \hline
\end{tabular} \vspace{1mm}
\caption{Number of samples used for training and evaluation}
\label{tab:samples}
\end{table}

\subsection{Training and results}

\begin{table}[h]
\centering
\begin{tabular}{|l|cc|}
\hline
Hyperparamaters            & \multicolumn{1}{c|}{Baseline} & Our solution          \\ \hline
Flow embedding size        & \multicolumn{2}{c|}{64}                               \\ \hline
Link embedding size        & \multicolumn{2}{c|}{64}                               \\ \hline
Device embedding size      & \multicolumn{1}{c|}{-}        & 16                    \\ \hline
MP max iterations          & \multicolumn{1}{c|}{8}        & 40                    \\ \hline
MP convergence threshold   & \multicolumn{1}{c|}{-}        & 0.05                  \\ \hline
Learning rate              & \multicolumn{1}{c|}{$10^-3$}  & $2.5 \times 10^-4$    \\ \hline
Number of training epochs* & \multicolumn{1}{c|}{98}         & \multicolumn{1}{c|}{86} \\ \hline
Loss function              & \multicolumn{1}{c|}{MAPE}     & log MSE               \\ \hline
\end{tabular} \vspace{1mm}
\caption{Hyperparameters of both models}
\label{tab:hyperparams}
\end{table}

We implemented both our solutions using Tensorflow 2.11. The baseline model's architecture is based on RouteNet-Fermi's architecture~\cite{ferriolgalmés2022routenetfermi} but with small changes to work in the testbed environment. Specifically, the message-passing component only considers links and flows, and the flow embeddings are defined using each flow's size in bits, its number of packets, the size of those packets, and its distribution (either CBR or MB) represented by two one-hot encoded features.

After iterating with multiple combinations of hyperparameters and several grid searches, we decided on the hyperparameter values present in table \ref{tab:hyperparams}. In the case of the number of epochs, both models were allowed to run with a redundant amount of epochs, and the best model was taken from the epoch in which the validation error was minimized (the table reflects at which epoch this occurred). Note the baseline model does not consider node embeddings, and only ends the message-passing process after a fixed number of iterations.

Another difference between both models is the loss function used: the original model uses the Mean Absolute Percentage Error (MAPE), as it is the metric we wish to minimize. The MAPE has the advantage of being a relative metric, making it independent of the scale of the true values. However, using it as the loss function introduces a bias where the model tends to underestimate its predictions \cite{MCKENZIE2011259}. As a result, our proposed solution is instead trained using the MSE of the logarithm of both the prediction and true value.

\begin{figure}[h]
     \centering
     \begin{subfigure}[b]{0.5\textwidth}
         \centering
         \includegraphics[width=\textwidth]{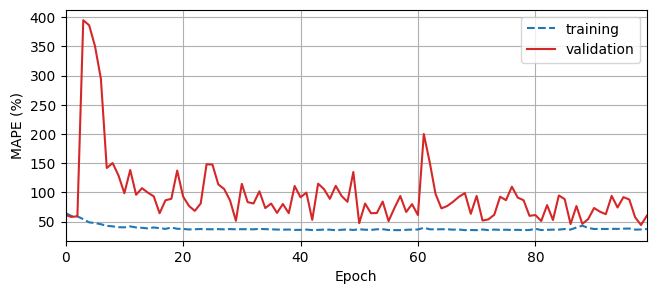}
         \caption{Baseline model}
         \label{fig:loss_baseline}
     \end{subfigure}
     \begin{subfigure}[b]{0.5\textwidth}
         \centering
         \includegraphics[width=\textwidth]{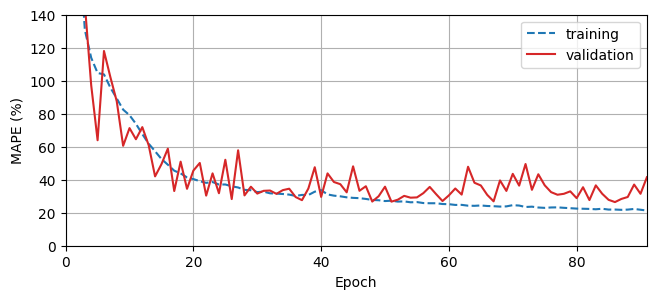}
         \caption{Our proposed solution}
         \label{fig:loss_solution}
     \end{subfigure}
        \caption{Loss evolution during training}
        \label{fig:loss_evolution}
\end{figure}

\begin{table}[h]
\centering
\begin{tabular}{|l|l|l|l|}
\hline
\textbf{Model}    & \textbf{Best epoch} & \textbf{Validation MAPE} & \textbf{Test MAPE} \\ \hline
Baseline          & 98                  & 44.298\%                 & \textit{100.528\%} \\ \hline
Proposed solution & 86                  & 26.449\%                 & \textit{27.831\%}  \\ \hline
\end{tabular} \vspace{1mm}
\caption{Test partition MAPE scores of both the baseline model and proposed solution}
\label{tab:final_mape}
\end{table}

\newpage The results of training both the baseline and our proposed solution can be examined in Figure \ref{fig:loss_evolution}. Specifically, Figure \ref{fig:loss_baseline} shows the training and validation MAPE as training evolves in the baseline model, while Figure \ref{fig:loss_solution} shows the evolution of the MAPE in the proposed solution. Overall, it is clear that the baseline model is inadequate to learn the complexities of the dataset only reaching a validation MAPE of 44\%, while our proposed solution does show the ability to learn and generalize, reaching a lower bound of 26.449\% MAPE. These results are further confirmed in Table \ref{tab:final_mape}, which shows how the best epoch from each model fares against the test dataset. While the baseline fails to generalize, and scores a MAPE barely over 100\% in the test dataset, the proposed solution correctly generalizes and gets a MAPE of 27.831\%.

\section{Related Work}
\label{sec:relatedwork}
Current state-of-the-art network models are varied in how they balance accuracy and computational cost. Overall, we might find that network models fall into one of three approaches: purely simulator-based, purely Machine Learning (ML) based models, and simulator-ML hybrids.

Network models based on simulators tend to be the most common, due to their high accuracy albeit at the cost of high simulation times, which increases exponentially to the scenario size. This includes popular simulators like ns-3~\cite{Riley2010} and OMNeT++~\cite{Varga2019}, both based on Discrete Event Simulation (DES). New simulators have been published, such as DONS~\cite{10.1145/3603269.3604844}, with improved implementations that decrease computation costs and allow for the simulation to be parallelized, but are still limited by the poor scaling to simulation size. On the other hand, trace-based simulation like Parsimon~\cite{285196} is faster than DES, but at the cost of lower accuracy.

Next, simulator-ML hybrid network models aim to replace part of the simulation with ML models. Doing so decreases simulation time, by avoiding to simulate segments of the network, while also opening the possibility of exploiting parallelization, another a major drawback of DES. However this comes at the cost of having a marginally worse accuracy. This hybrid approach was first popularized by MimicNet~\cite{10.1145/3452296.3472926} and it was later expanded and improved on by DeepQueueNet~\cite{10.1145/3544216.3544248}. While faster than pure simulation, these approaches still struggle to simulate larger network scenarios at a reasonable computational cost. Additionally, just as pure simulators, these solutions only consider idealized scenarios that may not reflect real network scenarios encountered by network administrators.

Finally, there exist pure ML network models, which are trained using simulators that generate its samples, but then do not require simulation during inference. This last property makes ML models' inference several degrees of magnitude faster than both simulation and hybrid models. While multiple NN architectures were tested \cite{10.1145/3152434.3152441, 10.1145/3229543.3229549}, eventually GNNs came out as the most appropriate architecture due to their ability to generalize across unseen network topologies. An example of these is the RouteNet family of models \cite{9109574, 9796944}, with RouteNet-Fermi being its latest iteration \cite{ferriolgalmés2022routenetfermi}. These models do not only infer quickly but are also adapted to deal with networks much larger than the ones seen during training. However, this comes at the cost of being less accurate than the other approaches. Also, as they are trained using simulated data, they also learn to only consider idealized scenarios and may expect certain information as input which in practice would be unknown to the network operators.

\section{Conclusion}
\label{sec:conclusion}
In this paper, we have presented a new GNN-based architecture for network modelling designed to be trained with captured traces rather than with simulated ones, as a possible solution to the Graph Neural Network Challenge 2023. In it, we have seen how we can encode packet traffic to better represent traffic distributions, and a new message passing algorithm that is designed to better model the relationships between traffic flows, physical links, devices and their ports.

Yet, there are aspects of the proposed solution that we wish to study further. For example, while we propose a method to encode packet sequences, many of the design decisions must be further studied. This includes how the sequences were summarized by aggregating information by milliseconds, or the use of a two-layered RNN to encode its information rather than any other sequential model like a bidirectional RNN or a transformer. Similarly, while in the Challenge we do not provide the packet traffic during the start of the simulation, as in theory this traffic represents an unstable and transient state of the network, it would be interesting to study a way of extracting useful information from it as well. Additionally, this Challenge we have simplified certain aspects of the problem, such as including scenarios with queuing policies, and will eventually need to be revisited. Ultimately, we have decided to leave all of these questions open for future work.

\bibliographystyle{ieeetr}
\bibliography{references}

\end{document}